\documentclass[aps,prl,twocolumn,superscriptaddress]{revtex4}
\usepackage{floatrow}
\usepackage{sidecap}
\usepackage{graphicx}
\usepackage{bm}
\usepackage{amssymb,amsmath}
\usepackage{upgreek}

\begin{document}

\title {Marginally Self-Averaging One-Dimensional Localization in Bilayer Graphene}

\author{Md.~Ali~Aamir}
\email{amohammed@iisc.ac.in}
\affiliation{Department of Physics, Indian Institute of Science, Bangalore 560 012, India.}
\author{Paritosh~Karnatak}
\email{paritosh@iisc.ac.in}
\affiliation{Department of Physics, Indian Institute of Science, Bangalore 560 012, India.}
\author{Aditya~Jayaraman}
\email{jaditya@iisc.ac.in}
\affiliation{Department of Physics, Indian Institute of Science, Bangalore 560 012, India.}
\author{T.~Phanindra~Sai}
\affiliation{Department of Physics, Indian Institute of Science, Bangalore 560 012, India.}
\author{T.~V.~Ramakrishnan}
\affiliation{Department of Physics, Indian Institute of Science, Bangalore 560 012, India.}
\author{Rajdeep~Sensarma}
\affiliation{Department of Theoretical Physics, Tata Institute of Fundamental Research, Dr. Homi Bhabha Road, Mumbai 400005, India.}
\author{Arindam~Ghosh}
\affiliation{Department of Physics, Indian Institute of Science, Bangalore 560 012, India.}
\affiliation{Centre for Nano Science and Engineering, Indian Institute of Science, Bangalore 560 012, India.}

\begin{abstract}
\textbf{The combination of field-tunable bandgap, topological edge states, and valleys in the band structure, makes insulating bilayer graphene a unique localized system, where the scaling laws of dimensionless conductance $g$ remain largely unexplored. Here we show that the relative fluctuations in $\ln g$ with the varying chemical potential, in strongly insulating bilayer graphene (BLG) decay nearly logarithmically for channel length up to $L/\xi \approx 20$, where $\xi$ is the localization length. This `marginal' self-averaging, and the corresponding dependence of $\langle\ln g\rangle$ on $L$, suggest that transport in strongly gapped BLG occurs along strictly one-dimensional channels, where $\xi \approx 0.5\pm0.1~\upmu$m was found to be much longer than that expected from the bulk bandgap. Our experiment reveals a nontrivial localization mechanism in gapped BLG, governed by transport along robust edge modes.}
\end{abstract}

\maketitle

The nature of sub-gap electrical transport in BLG at large transverse electric fields ($D$) has led to considerable debate~\cite{Oostinga2008,Zou2010,Taychatanapat2010,Yan2010,Allen2015,Zhu2017}. $D$ lifts the inter-layer symmetry, opening a bandgap in the quasi-particle energy spectrum. At large $D$, the charge carriers in BLG are strongly localized when the Fermi level is tuned close to the charge neutrality point (CNP)~\cite{Oostinga2008}. Moreover, gapped BLG behaves as a `marginal topological insulator', where the finite Berry phase and field-induced inversion symmetry breaking lead to topologically protected one-dimensional (1D) conduction modes along specific edge and stacking boundary configurations~\cite{Li2010,Vaezi2013,Zhang2013}. While this has been experimentally verified through observation of the valley Hall effect~\cite{Shimazaki2015a,Sui2015} and ballistic 1D channels along artificial~\cite{Li2016} and natural~\cite{Ju2015}  stacking boundaries, the topological properties also raise doubts on the current understanding of the localized state transport in gapped BLG at low temperature ($T < 50$~K). Although initial results were analyzed in terms of two-dimensional (2D) Mott-type variable range hopping (VRH) associated with localized states in the bulk~\cite{Oostinga2008,Taychatanapat2010,Zou2010,Miyazaki2010,Yan2010}, recent supercurrent interferometry experiments~\cite{Allen2015,Zhu2017} suggest strong edge-mode transport in short gapped BLG transistors. While this seems consistent with the apparent saturation of $g$ at large $D$ reported recently~\cite{Zhu2017}, the dimensionality of localized state transport in generic gapped BLG remains uncertain so far.

In BLG subjected to large $D$ at low $T$, the localization at the edge (due to short range lattice defects, chemical adsorbates etc.) and that in the gapped bulk are hard to distinguish because of limited experimental temperature range for VRH. Here, we have followed a new route based on evaluating the full conductance statistics in the insulating regime and we specifically study its self-averaging properties. A macroscopic variable $X$ in a disordered system of linear dimension $L$ is spatially ergodic, or self-averaging, when the relative fluctuations $R_X = \langle(\Delta X)^2\rangle/\langle X\rangle^2 \rightarrow 0$ as $L \rightarrow \infty$, where $\langle...\rangle$ represents averaging over different realizations of disorder. For strongly localized noninteracting carriers the electrical conductance $g$ (in units of $e^2/h$) does not self-average, but the logarithm of $g$ does~\cite{Anderson1980,abrikosov1981paradox,Beenakker1997,Muttalib1999}, in a manner that is uniquely sensitive to the dimensionality and the scaling properties of Anderson localization for $L \gg \xi$, the localization length~\cite{Lee1984,Serota1986,Raikh1989,Hughes1996,shi2014microwave,Aharony1996}. In two and three dimensions, the ensemble fluctuations in $\ln g$ are strongly self-averaging with $R_{\ln g} \sim L^{-d}$ ($d = 2, 3$)~\cite{Serota1986,Aharony1996}, whereas in 1D disordered systems at finite $T$, $\ln g$ is only marginally self-averaging because $R_{\ln g}$ decays logarithmically with $L$~\cite{Lee1984,Serota1986,Raikh1989}. This purely 1D effect, which so far remains experimentally elusive, to the best of our knowledge, is predicted to occur because the conductance of the system is determined primarily by the most resistive, but unavoidable, hop at the percolation threshold~\cite{Hughes1996}. To determine $R_{\ln g}$, we directly obtain the mean and the variance of $\ln g$ by measuring the full conductance probability distribution function (PDF) in the localized state for many dual gated BLG devices with varying channel lengths. We find that for small electric fields (typically $|D| \lesssim 0.5$~V/nm) the relative fluctuations in $\ln g$ with the Fermi level close to the CNP decay with $L$ as $\sim 1/L^2$, but the decay becomes nearly logarithmic at larger $D$ $-$ a characteristic of strictly 1D localized transport. 

\begin{figure*}
\includegraphics[width=1\linewidth]{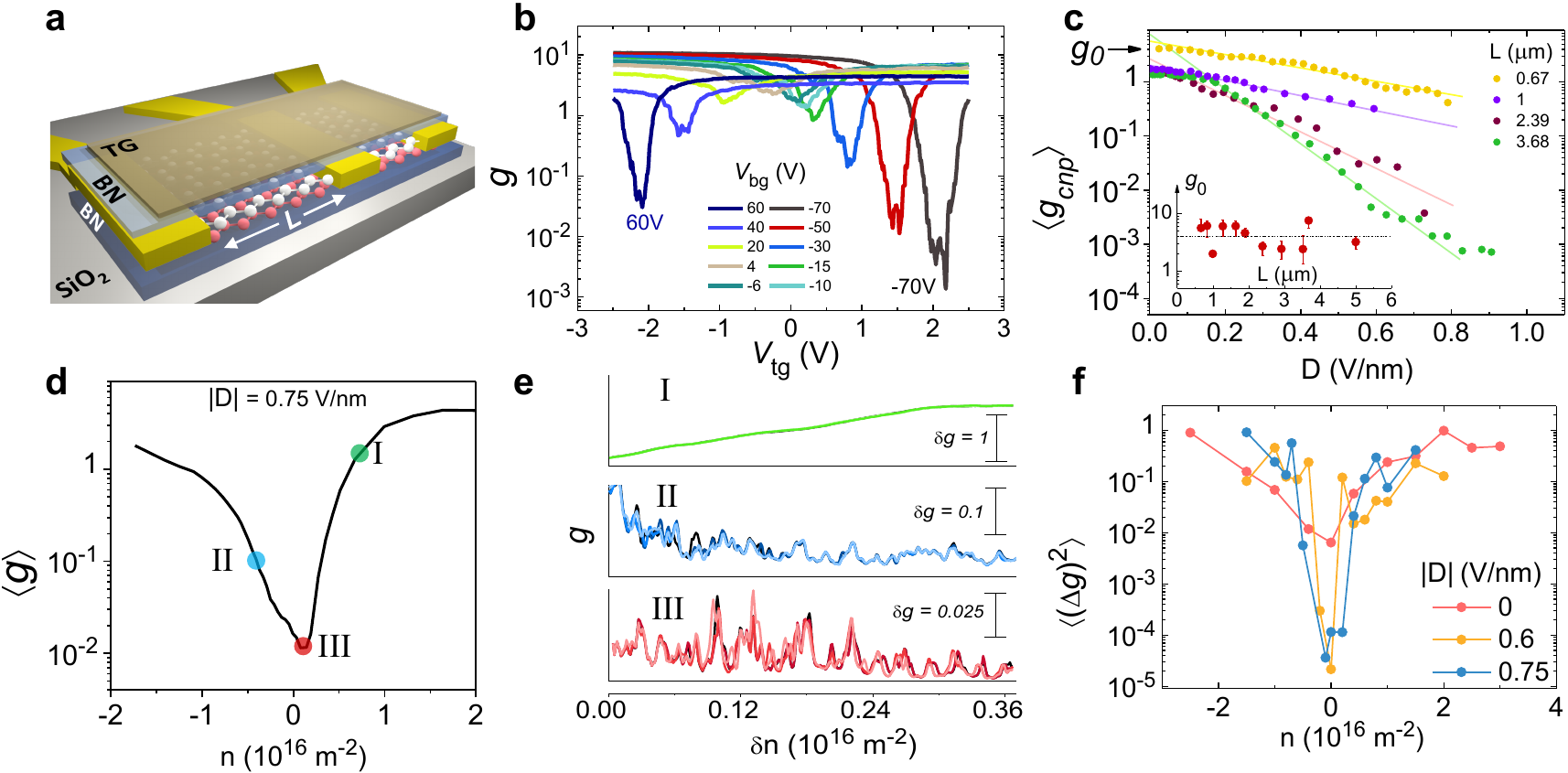}
\caption{\textbf{Device characteristics and conductance fluctuations.} \textbf{a.} Device schematic. \textbf{b.} Transfer characteristics ($g-V_{{\rm tg}}$) for a few $V_{{\rm bg}}$ at 275~mK. \textbf{c.} Mean conductance $\langle g_{{\rm cnp}} \rangle$ at the charge neutrality point (CNP) as a function of $D$ for several devices with different channel lengths. Inset shows extrapolated values of $\langle g_{{\rm cnp}} \rangle$  to $D=0$~V/nm. \textbf{d.} Carrier density ($n$) dependence of $\langle g \rangle$ for $|D|=0.75$~V/nm. \textbf{e.} Conductance fluctuations as a function of small variations in $n$ induced by the top gate at three different points marked in \textbf{d}. \textbf{f.}~Variance of conductance fluctuations, $\langle (\Delta g)^2 \rangle$, as a function of $n$ for several $D$.}
\end{figure*}
\label{Fig1}

The dual-gated BLG channels were created with mechanical exfoliation, followed by either one (top) or both sides covered with hexagonal boron nitride (hBN). Both surface and edge contacting methods were adopted~\cite{Karnatak2016}, and the top gate length defines the channel length $L$. Representative schematic of an edge-contacted device is shown in Fig.~1a, and more details can be found in Ref.~\cite{Suppl}. We have studied the conductance statistics of 20 BLG channels with $L$ ranging from $\sim 0.7 - 19.5~\upmu$m in different BLG flakes with channel widths ranging from $\sim 1.0 - 3.1~\upmu$m. Although measurements were carried out down to $\sim20$~mK, the conductance becomes nearly insensitive to $T$ below $\lesssim 1$~K in all devices in the localized regime (Fig.~3c). Fig.~1b illustrates the typical transfer characteristics of gated BLG, obtained at $T=275$~mK in device Dev3 with $L = 1.28~\upmu$m, where the two-probe conductance $g$ is shown as a function of the top gate voltage ($V_{{\rm tg}}$) at fixed backgate voltages ($V_{{\rm bg}}$). 

Fig.~1b also shows that the conductance at CNP, $g_{{\rm cnp}}\ll 1$ for large values of $|V_{{\rm bg}}|$, which implies strong localization of carriers at the center of the bandgap as $|D|$ increases. Between $|D| = 0$ and $\lesssim 0.4$~V/nm, when localization in the bulk is weak, the variation in $g_{{\rm cnp}}$ with $D$ is device dependent. However, for $|D| > 0.5$~V/nm, $g_{{\rm cnp}}$ decreases nearly exponentially in almost all our devices irrespective of $L$ (Fig.~1c). Absence of saturation in $g_{{\rm cnp}}$ at large $D$ confirms that there are no accidental stacking/grain boundaries that shunt the source and drain leads~\cite{Ju2015}, and that $L>\xi$ in all devices. To avoid gate leakage, the maximum $|D|$ was limited to $\approx 0.9 - 1.3$~V/nm which, for longer devices, led to $g_{{\rm cnp}}$ as low as $\sim 10^{-3} - 10^{-4}$.

The key feature of the transfer characteristics in Fig.~1b is the strong relative fluctuations in $g$ in the vicinity of the CNP, which become more apparent as $|D|$ increases. The fluctuations are remarkably reproducible as the Fermi level is varied, as illustrated in Fig.~1e with device Dev6 ($L = 2.92~\upmu$m) at $T = 0.03$~K and $|D| = 0.75$~V/nm. The panels of Fig.~1e represent three separate regimes from metallic (large $n$) to strongly localized (CNP, $n \approx 0\times10^{16}$~m$^{-2}$) transport, where four traces of $g$ within the window $\delta n = 0.36\times10^{16}$~m$^{-2}$ demonstrate the reproducibility, and confirm negligible contribution from time-dependent noise~\cite{PalPRL2009,Karnatak2016}. Within $\delta n$, which is similar to the spatial variation in carrier density typically present in BLG~\cite{Rutter2011}, the fluctuations reflect traversing across the microscopic realizations at a fixed point in the phase space~\cite{Lee1984,Serota1986,Hughes1996}. 

\begin{figure*}
\includegraphics[width=1\linewidth]{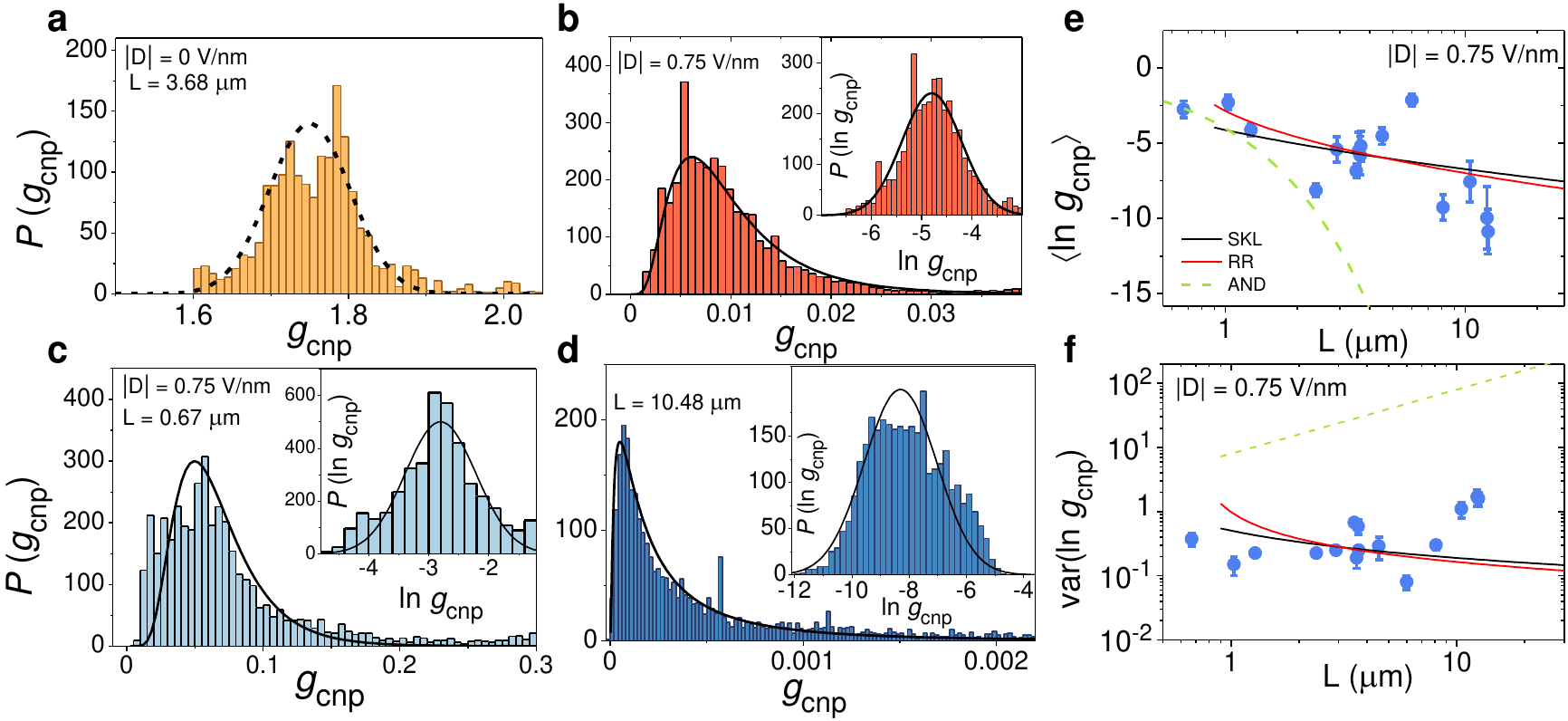}
\caption{\textbf{Conductance distribution.} Probability distribution function of conductance ($P(g_{{\rm cnp}})$ and $P(\ln g_{{\rm cnp}})$) around the CNP for different L and D, with a fixed length $L=3.68$~$\upmu$m at \textbf{a.} $D=0$~V/nm and \textbf{b.} $|D|=0.75$~V/nm; \textbf{c} and \textbf{d}  for a fixed $|D|=0.75$~V/nm at \textbf{c.} $L=0.67~\upmu$m and \textbf{d.}~$L=10.48~\upmu$m. Insets in \textbf{b-d} show the corresponding probability distributions of $\ln g_{{\rm cnp}}$. The dashed and solid lines represent normal and log-normal distributions respectively. \textbf{e.}~Average logarithm of conductance around the CNP ($\langle \ln g_{{\rm cnp}} \rangle$) as a function of $L$ for $|D|=0.75$~V/nm. The expected variations of $\langle \ln g_{{\rm cnp}} \rangle$ from SKL~\cite{Serota1986}  (black solid line), RR~\cite{Raikh1989} (red solid line) and Anderson model~\cite{Beenakker1997} (green dashed line) are also indicated. The error bars indicate the standard deviation of $\langle \ln g_{{\rm cnp}} \rangle$ over ensemble variations. \textbf{f.}~The variance of $\ln g_{{\rm cnp}}$, ${\rm var}(\ln g_{{\rm cnp}})$ as a function of channel length $L$ for $|D|=0.75$~V/nm.  The $L$ $-$ dependence of variance expected from SKL, RR and Anderson model, with $\xi\approx0.5~\upmu$m are shown. The error bars represent the uncertainty in log-normal fits.}
\end{figure*}
\label{Fig2}

To quantify, we have first calculated the variance $\langle(\Delta g)^2\rangle$ within $\delta n$ (consisting of $\approx 400$ points or realizations), and shown it as a function of $n$ in Fig.~1f for three values of $D$. At large $n$ (typically $|n|> 1\times 10^{16}$~m$^{-2}$), the onset of quasi-metallic or weakly localized regime is characterized by $g \gtrsim 1$, where $\langle(\Delta g)^2\rangle$ saturates to $\approx 0.1 - 1$, irrespective of $D$. This is universal conductance fluctuations due to quantum interference of multiply backscattered electron waves~\cite{Lee1985,Pal2012}. Since the Fermi level lies within the conduction or valence bands, this is a bulk phenomenon, expected for diffusive 2D disordered systems when the phase coherence length is similar to $L$. As $|n|$ decreases, $\langle(\Delta g)^2\rangle$ decreases, exhibiting a minimum around the CNP. The reduction in $\langle(\Delta g)^2\rangle$ at CNP is weak for $D = 0$~V/nm, but is nearly five orders of magnitude for $|D| \geq 0.6$~V/nm. The fluctuations in the localized regime are largely immune to contact effects~\cite{Karnatak2016}, where the channel conductance is at least an order of magnitude lower than that of the contacts.

\begin{figure}
\includegraphics[width=1\linewidth]{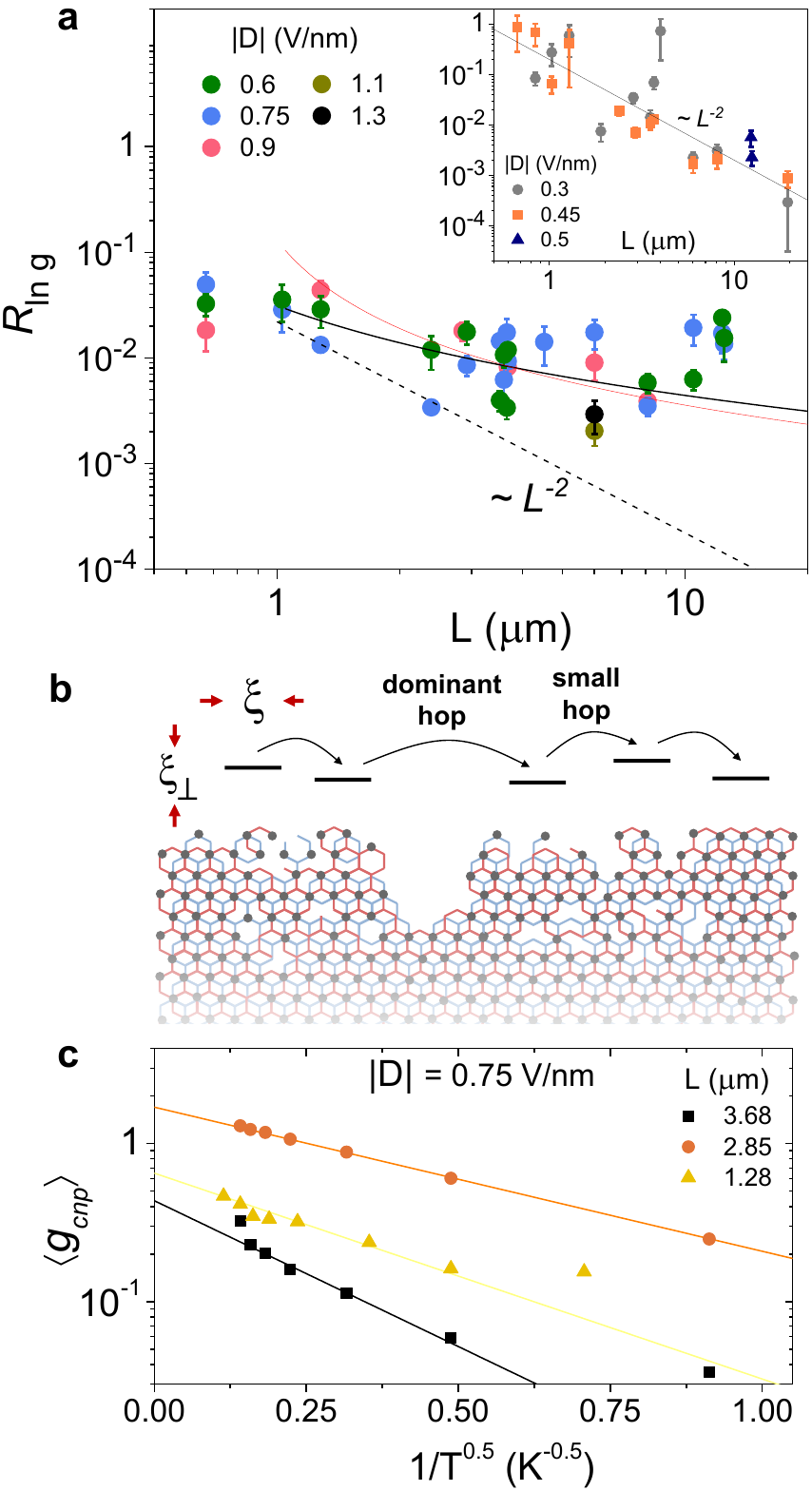}
\caption{\textbf{Self-averaging of $\ln g_{{\rm cnp}}$ in gapped bilayer graphene.} \textbf{a.}~Relative fluctuations $R_{\ln g}={\rm var}(\ln g_{{\rm cnp}})/\langle \ln g_{{\rm cnp}} \rangle^{2}$ as a function of $L$ for high $|D|$~($>0.5$~V/nm). The solid lines represent a logarithmic decay $R_{\ln g}\propto\left[\ln\left(2L/\xi\right)\right]^{-2}$ expected from the SKL and RR model with $\xi\approx0.5~\upmu$m. Inset shows $R_{\ln g}$ vs $L$ for $|D|=0.3$, $0.45$ and $0.5$~V/nm. Here, strong self-averaging is indicated by the decay in $R_{\ln g}$ as $R_{\ln g} \propto L^{-2}$ (solid line). The error bars have been computed as the net error from uncorrelated relative errors in $\langle \ln g_{{\rm cnp}} \rangle$ and ${\rm var}(\ln g_{{\rm cnp}})$. \textbf{b.}~Schematic depicting hopping transport along disordered 1D edge of the bilayer graphene with gapped bulk. \textbf{c.}~$\langle g_{\rm cnp} \rangle$ as a function of $T^{-0.5}$ for 3 devices at a fixed $|D|=0.75$~V/nm, where the fits $\langle g \rangle \sim \exp -(T_0/T)^{1/2}$ indicate the validity of 1D hopping transport.}
\end{figure}
\label{Fig3}

Fluctuations in $g$ with gate voltage in the strongly localized regime may be caused by inhomogeneous charge distribution or single-particle localized states in the bulk of the BLG~\cite{Liao2010} through, {\it e.g.}, local charging of electron/hole puddles~\cite{beenakker1991quantum} or multiple transmission resonances~\cite{Azbel1984}. However, direct charging or resonance effects are expected to decay rapidly for $L \gtrsim O[100~{\rm nm}]$, the typical scale of inhomogeneity. Such fluctuations may also arise in disordered quasi-1D or (short) 2D systems due to the extreme sensitivity of the critical resistance-determining hop to the local chemical potential~\cite{Ambegaokar1971}, as shown for short conducting channels in silicon and semiconductor heterostructures~\cite{Fowler1986,Hughes1996,hartstein1984one}. This mechanism manifests in the probability distribution function (PDF) of $\ln g$ that approaches a Gaussian for large $L$~\cite{Serota1986,Raikh1989,Beenakker1997,shi2014microwave}, and the characteristic $L$-dependence of the mean and variance of $\ln g$ depend sensitively on the dimensionality of the system~\cite{Lee1984,Serota1986,Raikh1989}. 

Fig.~2a-d show the PDF of the fluctuations in $g$ observed within the $\delta n$ window around the CNP for different values of $D$ and $L$. Fig.~2a and 2b present data from device Dev10 with $L = 3.68~\upmu$m at different $D$. At $D = 0$~V/nm (Fig.~2a), the device is quasi-metallic ($g_{{\rm cnp}} \simeq  2$), and PDF of $g_{{\rm cnp}}$ is close to a Gaussian, symmetric around $\langle g_{{\rm cnp}}\rangle$ (dashed line). However, at large $|D|$ of 0.75~V/nm (Fig.~2b), the device is strongly localized at the CNP with modal $g_{{\rm cnp}} \sim 0.01$, and PDF in $g_{{\rm cnp}}$ is strongly asymmetric around the peak. Instead, as shown in the inset of Fig.~2b, the PDF of $\ln g_{{\rm cnp}}$ is symmetric around $\langle\ln g_{{\rm cnp}}\rangle$, and corresponds closely to a Gaussian distribution (solid line). The log-normal PDF in $g_{{\rm cnp}}$ is observed for all but one $L$ at large $D$ (typically $|D|\gtrsim0.5$~V/nm, see Ref.~\cite{Suppl} for details), as illustrated with two other devices in Fig.~2c and 2d. Occasionally, the distribution can exhibit weak asymmetry due to blocking effect or ``optimal shorts or punctures'' in long and short channels, respectively~\cite{Raikh1989,Hughes1996} (Fig.~S3).

Log-normal conductance PDF in strongly localized systems, when  $L \gg \xi$, has been analytically shown in 1D systems~\cite{Muttalib1999,Beenakker1997} whereas only numerically in higher dimensions~\cite{Cohen1988,Aharony1996}. The Gaussian fits to the PDFs allow a direct evaluation of $\langle\ln g_{{\rm cnp}}\rangle$ and ${\rm var}(\ln g_{{\rm cnp}})$ with varying $D$ and $L$, shown in Fig.~2e and 2f respectively. Quantitatively, the mean conductance of a 1D disordered system can be expressed as~\cite{Serota1986,Raikh1989}, 

\begin{equation}
\langle\ln g\rangle \approx -\left(\frac{T_0}{T}\right)^{1/2} f(L/\xi, T_0/T)
\end{equation}
\label{mean}
    
\noindent where, $T_0 = 1/k_B\xi N$, and $N$ is the 1D density-of-states at the Fermi level. The functional form of $f$ depends on the details of the hopping mechanism. Serota, Kalia and Lee~\cite{Serota1986} (SKL) considered that selective links which are the weakest dominate the net conductance, and showed that $f \approx [\ln(2L/\xi)]^{1/2}$. On the other hand, Raikh and Ruzin~\cite{Raikh1989} (RR) introduced the concept of an optimal break in the phase space of energy and position of the localized states, and derived $f \approx [\ln[(L/\xi)(T/T_0)^{1/2}\ln^{1/2}(L/\xi)]]^{1/2}$. Moreover, the variance of conductance for a 1D disordered system is predicted to exhibit a logarithmic decrease ($\sim\left[\ln\left(2L/\xi\right)\right]^{-1}$)~\cite{Serota1986,Raikh1989}. Therefore the relative fluctuations, obtained by dividing ${\rm var}(\ln g_{{\rm cnp}})$ with the corresponding $\langle\ln g_{{\rm cnp}}\rangle^{2}$, are expected to only marginally decay with $L$ as $R_{\ln g}\sim\left[\ln\left(2L/\xi\right)\right]^{-2}$ in 1D.

Fig.~3a shows the $L$-dependence of the relative fluctuations $R_{\ln g}$ in the strongly localized regime ($|D| \gtrsim 0.5$~V/nm). Clearly, $R_{\ln g}$ remains nearly constant or decreases marginally even as $L$ increases by more than an order of magnitude. This represents near absence of self-averaging in the localized BLG transport. While SKL and RR~\cite{Serota1986,Raikh1989} generally capture the logarithmic decay in self-averaging with $L$ (solid lines in Fig.~3a), we emphasize that the marginal self-averaging behavior is a model-independent phenomenon. This is expected in a randomly disordered purely 1D system because the most resistive link cannot be bypassed with an increasing system size. Intriguingly, best fit to the data, both by SKL and RR, yields a similar estimate of $\xi \approx 0.5~\upmu$m. This estimate exceeds the localization length $\sim \hbar/\sqrt{2m^*\Delta} \sim 3 - 10$~nm due to the bulk bandgap ($\Delta$) by nearly two orders, but is similar to the length scale of short-range edge defects and inter-valley scattering~\cite{Gorbachev2007}, suggesting transport along the BLG edge~\cite{Zhu2017}. The insensitivity of $R_{\ln g}$ to the magnitude of $D$ for $|D| \gtrsim 0.5$~V/nm, further suggests that beyond a certain gap, the bulk of the BLG becomes largely inconsequential to the hopping transport. Importantly, the self-averaging properties expected in two dimensions are recovered at low $|D|$, where bulk transport contributes significantly, and manifest in $R_{\ln g} \sim 1/L^2$~\cite{Serota1986,Aharony1996} for $|D| \lesssim 0.5$~V/nm (inset of Fig.~3a). 

The $\langle\ln g_{{\rm cnp}}\rangle^{2}$ and the ${\rm var}(\ln g_{{\rm cnp}})$, shown in Fig.~2e and 2f respectively, can also be roughly described by the SKL and RR models with the same  $\xi \approx 0.5 \pm 0.1~\upmu$m as used above but are inconsistent with the form expected from the $T=0$~K Anderson localization model~\cite{Beenakker1997}. Importantly, both SKL and RR models suggest $\langle\ln g\rangle \sim -(T_0/T)^{1/2}$ in the leading order, which is indeed observed in our experiments in the range $T \gtrsim 1$~K as shown in Fig.~3c. 

%Goes to suppl.
%This dependence, however, seems to saturate for low temperatures $T \lesssim 1$~K possibly because of insufficient electron cooling due to weak electron - phonon coupling in graphene, at the minimum experimental source drain bias of $\sim 100~\upmu$V (see Supplementary for details).%

%, corresponding to $L/\xi$ as large as $\approx 20$

%The variance ${\rm var}(\ln g_{{\rm cnp}})=\langle (\ln g_{{\rm cnp}} - \langle \ln g_{{\rm cnp}} \rangle)^2 \rangle$, obtained from the PDF, is shown in Fig.~3b as a function of $L$ for $|D|=0.75$~V/nm. The observed behavior of ${\rm var}(\ln g_{{\rm cnp}})$ with $L$ is inconsistent with a logarithmic decrease ($\sim\left[\ln\left(2L/\xi\right)\right]^{-1}$) expected from SKL or RR~\cite{Serota1986,Raikh1989}, or even a $\propto L/\xi$ dependence expected from the $T=0$~K Anderson localization model~\cite{Beenakker1997}, for $\xi \approx 0.5~\upmu$m (shown in Fig.~3b). Note that, the increasing trend of ${\rm var}(\ln g_{{\rm cnp}})$, which is also inconsistent with the 2D hopping scenario~\cite{Serota1986,Hughes1996}, is not a finite size effect~\cite{shi2014microwave} because it continues even when $L/\xi \sim 20$ (for the longest device Dev14). %

%Similar to ${\rm var}(\ln g_{{\rm cnp}})$, the magnitude of $R_{\ln g}$ is also insensitive to $|D|$ in this high field regime, suggesting that the marginal self-averaging is not directly linked to the bulk localized states.%

Alternative edge-bound transport processes, in particular, those due to lateral confinement at the BLG boundary~\cite{Allen2015} cannot be completely ruled out. However, the inherent tendency to bypass around strong disorder renders a quasi 1D nature to these channels and it is unlikely that such a mechanism would lead to the suppression of self averaging, which is a strictly 1D phenomenon. The self-averaging may also be absent in specific cases of fractal disorder landscape~\cite{Bunde1995} or proximity to critical point~\cite{Aharony1996}, which are not likely in the BLG devices. Thus in view of the theoretical~\cite{Li2010} and recent experimental reports~\cite{Allen2015,Zhu2017}, a likely mechanism is hopping via low-energy electronic states of disordered edges in BLG (see Fig.~3b). Notably, the observed pre-factor $g_0 \sim 2 - 6$ in the exponential variation of $g_{{\rm cnp}} = g_0\exp(-\alpha|D|)$ for $|D|\gtrsim 0.4$~V/nm (Fig.~1c inset), is similar to that expected in the localized edge transport~\cite{Li2010}, although the manner by which $D$ modifies the tunnelling probability of charge between two adjacent fragments needs further understanding (here $\alpha$ is a device-dependent parameter). 

Finally, we note an intriguing feature in the $T$-dependence of conductance in the strongly localized phase ($|D| = 0.75$~V/nm) as shown in Fig.~3c, where $\ln g_{\rm cnp} (\sim -(T_0/T)^{1/2})$ extrapolates to a pre-factor of the order of the conductance quantum ($\sim 0.5 e^2/h - 2 e^2/h$) in three separate channels. While 1D localized channels may naturally exhibit this in the $T \rightarrow \infty$ limit~\cite{Nazarov2003}, universal pre-factor in localized 2D electron systems has previously been attributed to electron-electron interaction-driven hopping transition~\cite{Khondaker1999, Ghosh2002}. In localized 1D systems ~\cite{Shepelyansky1994}, the effect of electron-electron interaction on hopping mechanism remains poorly understood, compounded by the difficulty in distinguishing the Efros-Shklovskii mechanism~\cite{Efros1975} with $\ln g \sim -(T_0/T)^{1/2}$, from the Mott hopping law. Nonetheless, recent spectroscopy~\cite{Magda2014} and transport~\cite{Kinikar2017} experiments reveal strong on-site electron-electron interaction along the edges of graphene.

In summary, our experiment probes self-averaging of the logarithm of conductance in strongly localized bilayer graphene with full conductance statistics as a function of the device length and bandgap. We observed a logarithmically slow marginal self-averaging, which is a strictly one-dimensional phenomenon, and may be connected to an interplay of topological states at the bilayer graphene edge and frozen disorder (edge lattice defects). 

\begin{acknowledgments}
We thank Michael Pepper, Subroto Mukerjee and Sumilan Banerjee for valuable discussions. The authors thank CSIR and DST for financial support.
\end{acknowledgments}

M. A. A., P. K. and A. J. contributed equally to this work.

\end{document}

\title{Supplemental Material: Marginally self-averaging one-dimensional localization in bilayer graphene}

\author{Md. Ali Aamir}
\email{amohammed@iisc.ac.in}
\affiliation{Department of Physics, Indian Institute of Science, Bangalore 560 012, India.}
\author{Paritosh~Karnatak}
\email{paritosh@iisc.ac.in}
\affiliation{Department of Physics, Indian Institute of Science, Bangalore 560 012, India.}
\author{Aditya~Jayaraman}
\email{jaditya@iisc.ac.in}
\affiliation{Department of Physics, Indian Institute of Science, Bangalore 560 012, India.}
\author{T.~Phanindra~Sai}
\affiliation{Department of Physics, Indian Institute of Science, Bangalore 560 012, India.}
\author{T. V. Ramakrishnan}
\affiliation{Department of Physics, Indian Institute of Science, Bangalore 560 012, India.}
\author{Rajdeep Sensarma}
\affiliation{Department of Theoretical Physics, Tata Institute of Fundamental Research, Dr. Homi Bhabha Road, Mumbai 400005, India.}
\author{Arindam~Ghosh}
\affiliation{Department of Physics, Indian Institute of Science, Bangalore 560 012, India.}
\affiliation{Centre for Nano Science and Engineering, Indian Institute of Science, Bangalore 560 012, India.}

%\author{M. A. Aamir}
%\email{aamir@physics.iisc.ernet.in}
%\affiliation{Department of Physics, Indian Institute of Science, Bangalore 560 012, India.}
%\author{Paritosh~Karnatak}
%\affiliation{Department of Physics, Indian Institute of Science, Bangalore 560 012, India.}
%\author{T.~Phanindra~Sai}
%\affiliation{Department of Physics, Indian Institute of Science, Bangalore 560 012, India.}
%\author{Arindam~Ghosh}
%\affiliation{Department of Physics, Indian Institute of Science, Bangalore 560 012, India.}
%\affiliation{Centre for Nano Science and Engineering, Indian Institute of Science, Bangalore 560 012, India.}

\maketitle

\renewcommand{\thefigure}{S\arabic{figure}}

\subsection{Device details}

Table below has the details of all the 20 devices we have used in our experiments. The channel length ranges from $0.67~\upmu$m to $19.51~\upmu$m, varied over an order of magnitude. The channel widths vary over a range from 1$~\upmu$m to 3.1$~\upmu$m. The mobility range is from  $~1500$~cm$^2$/Vs to $~16000$~cm$^2$/Vs.

\begin{center}
\title{Table I: Details of all the devices used in this work}
\begin{tabular}{|c|c|c|c|c|c|}
\hline
Device name & Channel Length & Channel Width & Mobility (cm$^2$/V s) & Substrate & Contacts type\\
\hline

%Dev1  & 0.67  & 1.26  & 1500  & SiO$_{2}$  & surface contacted \\
%Dev18  & 0.84  & 1  & $\sim$2664  & SiO$_{2}$  & surface contacted \\
%Dev2  & 1.03  & 1.75  & 5500  & BN  & edge contacted \\
%Dev3  & 1.28  & 1.26  & 1700  & SiO$_{2}$  & surface contacted \\
%Dev19  & 1.9  & 1.45  & $\sim$437  & SiO$_{2}$  & surface contacted \\
%Dev4  & 2.39  & 1.75  & 5800  & BN  & edge contacted \\
%Dev5  & 2.85  & 1.26  & 1950  & SiO$_{2}$  & surface contacted \\
%Dev6  & 2.92  & 1.75  & 4600  & BN  & edge contacted \\
%Dev7  & 3.52  & 1.75  & 5000  & BN  & edge contacted \\
%Dev8  & 3.6  & 1.7  & $\sim$16000  & BN  & edge contacted, Hall bar \\
%Dev9  & 3.66  & 1.7  & $\sim$16000  & BN  & edge contacted, Hall bar \\
%Dev10  & 3.68  & 1.26  & 3700  & SiO$_{2}$  & surface contacted \\
%Dev20  & 4  & 2.29  & $\sim$4293  & SiO$_{2}$  & surface contacted \\
%Dev11  & 4.52  & 1.7  & 16000  & BN  & edge contacted, Hall bar \\
%Dev15  & 6.0  & 3.1  & 15900  & BN  & edge contacted \\
%Dev16  & 8.1  & 1.3  & 10600  & BN  & edge contacted \\
%Dev12  & 10.48  & 1.7  & $\sim$16000  & BN  & edge contacted, Hall bar \\
%Dev13  & 12.34  & 1.7  & $\sim$16000  & BN  & edge contacted, Hall bar \\
%Dev14  & 12.48  & 1.7  & $\sim$16000  & BN  & edge contacted, Hall bar \\
%Dev17  & 19.51  & 3.0  & $\sim$10000  & BN  & edge contacted, Hall bar \\

Dev1 & 0.67 & 1.26 & 1500 & SiO$_2$ & surface contacted \\
Dev2 & 1.03 & 1.75 & 5500 & BN & edge contacted \\
Dev3 & 1.28 & 1.26 & 1700 & SiO$_2$ & surface contacted \\
Dev4 & 2.39 & 1.75 & 5800 & BN & edge contacted \\
Dev5 & 2.85 & 1.26 & 1950 & SiO$_2$ & surface contacted \\
Dev6 & 2.92 & 1.75 & 4600 & BN & edge contacted \\
Dev7 & 3.52 & 1.75 & 5000 & BN & edge contacted \\
Dev8 & 3.6 & 1.7 & $\sim$16000 & BN & edge contacted, Hall bar \\
Dev9 & 3.66 & 1.7 & $\sim$16000 & BN & edge contacted, Hall bar \\
Dev10 & 3.68 & 1.26 & 3700 & SiO$_2$ & surface contacted \\
Dev11 & 4.52 & 1.7 & 16000 & BN & edge contacted, Hall bar \\
Dev12 & 10.48 & 1.7 & $\sim$16000 & BN & edge contacted, Hall bar \\
Dev13 & 12.34 & 1.7 & $\sim$16000 & BN & edge contacted, Hall bar \\
Dev14 & 12.48 & 1.7 & $\sim$16000 & BN & edge contacted, Hall bar \\
Dev15 & 6.0 & 3.1 & 15900 & BN & edge contacted \\
Dev16 & 8.1 & 1.3 & 10600 & BN & edge contacted \\
Dev17 & 19.51 & 3.0 & $\sim$10000 & BN & edge contacted, Hall bar \\
Dev18 & 0.84 & 1 & $\sim$2660 & SiO$_2$ & surface contacted \\
Dev19 & 1.9 & 1.45 & $\sim$440 & SiO$_2$ & surface contacted \\
Dev20 & 4 & 2.29 & $\sim$4300 & SiO$_2$ & surface contacted \\

\hline

\end{tabular}
\end{center}
\label{Table1}

\newpage

\subsection{Independent control of carrier density and perpendicular electric field using dual-gated device design}

\begin{figure*}
\includegraphics[width=0.8\linewidth]{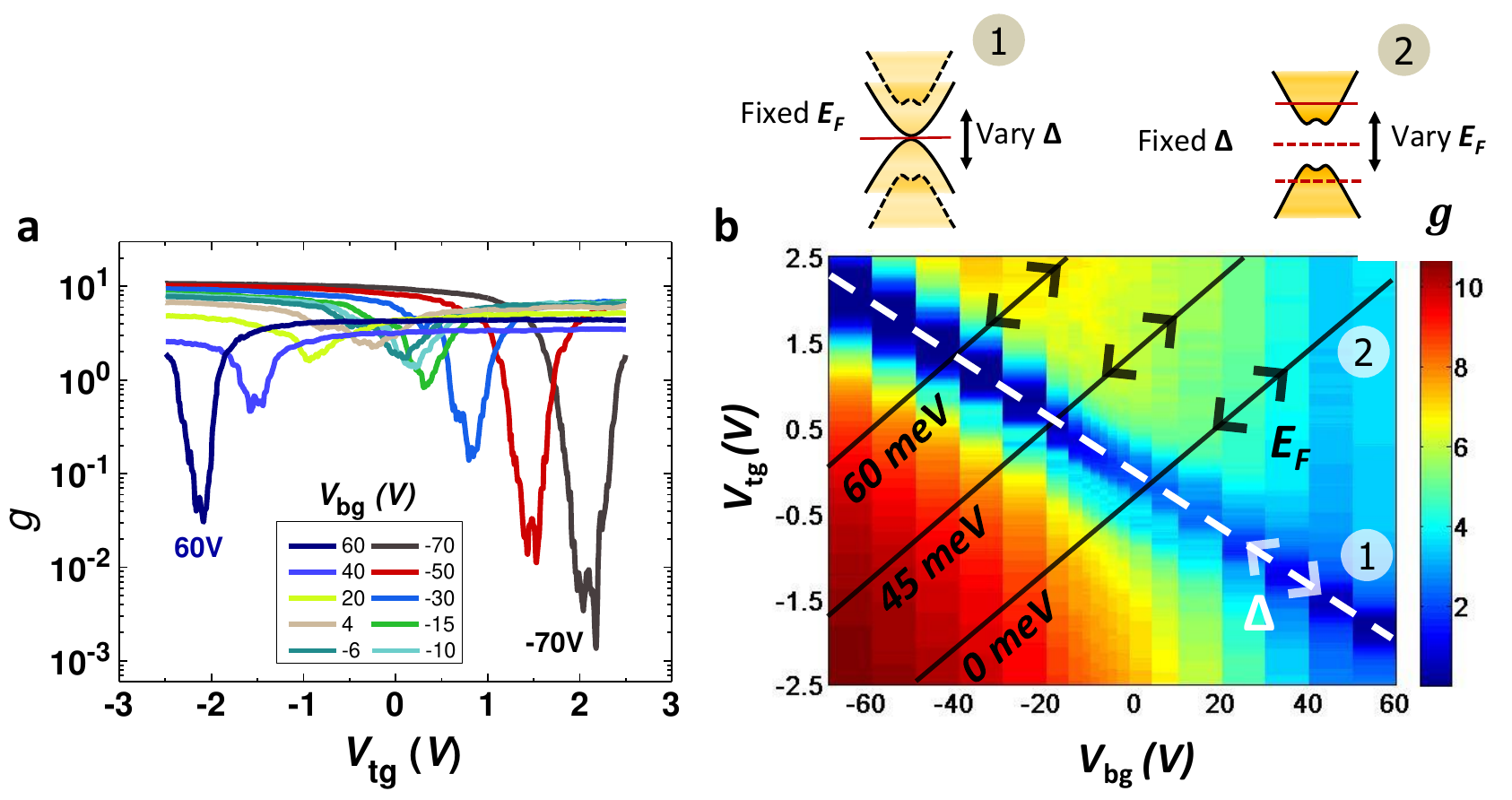}
\caption{\textbf{a.} Gate voltage characteristics of the conductance of the device, $g$, reproduced from Fig.~1b of the main text \textbf{b.} 2D colour plot of $g$ as a function of $V_{\rm bg}$ and $V_{\rm tg}$; loci of $(V_{\rm bg},V_{\rm tg})$ for fixed Fermi energy and varying band gap is labelled as (1) and its vice versa labelled as (2) with schematics of the corresponding process at the top of the graphs. Loci for $D=0, -0.45,$ and $-0.6$~V/nm are shown by black lines whereas $n=0$ m$^{-2}$ has been shown by dashed white lines.}
\end{figure*}
\label{FigS1}

Each of the two gates in a dual-gated BLG FET (as shown in Fig.~1a of the main text) contributes independently to the carrier density (as $n=C(V_G-V_0)/e$) and electric field (as $D=C(V_g-V_0 )/\epsilon_0$) where $C$ is the capacitance per unit area, $d$ being the dielectric thickness, $V_G$ is the applied gate voltage and $V_0$ is the gate voltage offset necessary to counterbalance any extrinsic doping.  The combined carrier density $n$ and electric field $D$ is then given by,

\[D \mathrm{(V/nm)}=\left[C_{\rm bg} (V_{\rm bg}-V_{\rm b0})-C_{\rm tg} (V_{\rm tg}-V_{\rm t0}) \right]/2\epsilon_0 \]
\[n (\mathrm{m}^{-2} )=\left[C_{\rm bg} (V_{\rm bg}-V_{\rm b0} )+C_{\rm tg} (V_{\rm tg}-V_{\rm t0})\right]/e \]

On fixing one gate, a background carrier density and band gap is fixed which can be tuned continuously using the other gate. Fig.~1b in the main text shows the gate voltage characteristics of the conductance $g$, which has been reproduced in Fig.~S1a. The back gate voltage $V_{bg}$ is fixed at several values and the conductance of this device is measured as the top gate voltage $V_{\rm tg}$ is continuously varied. At points where the net carrier density $n$ is tuned to zero, called the charge neutrality points (CNP), the conductance $g$ takes a minimum value, which we call $g_{\rm cnp}$. As the back gate voltage is changed, its contribution to carrier density is altered and a different $V_{\rm tg}$ is required to neutralise the carrier density to zero. That is why the position of the conductance minimum varies with $V_{\rm bg}$. Since induced carrier density is proportional to change in the gate voltages ($\Delta V_{\rm bg}$, $\Delta V_{\rm tg}$) with the corresponding capacitances ($C_{\rm bg}$, $C_{\rm tg}$) as the proportionality factor, the required change in $V_{\rm tg}$ to maintain constant carrier density for a given change in $V_{\rm bg}$ is given by

\[ \Delta V_{\rm tg} = \frac{C_{\rm bg}}{C_{\rm tg}}\Delta V_{\rm bg}\]

$g_{\rm cnp}$ also increases as we go further right or left with respect the central trace, because the band gap becomes wider due to increasingly larger perpendicular electric field across the bilayer graphene. The increase is roughly exponential in the magnitude of $D$. The central trace for which $g_{\rm cnp}$ is maximum has both the carrier density and band gap closest to zero values, $n=0$~m$^{-2}$, $D=0$~V/nm. The gate voltages at this point $V_{\rm b0}$, $V_{\rm t0}$ are residual voltages that have counterbalanced residual environmental doping and electric field.

The 2D colour plot of conductance as a function of both $V_{\rm bg}$ and $V_{\rm tg}$ is shown in Fig.~S1b. The set of ($V_{\rm bg}$,$V_{\rm tg}$) co-ordinates corresponding to CNP, i.e. for zero carrier density but varying electric field, is along the dashed white line running diagonally in the 2D colour plot, and also denoted by label (2). This is a straight line which can be derived by solving the above equations for $n=0$~m$^{-2}$:

\[V_{\rm tg}=-\frac{C_{\rm bg}}{C_{\rm tg}}  V_{\rm bg}+\frac{C_{\rm bg}}{C_{\rm tg}}  V_{\rm b0}+V_{\rm t0} \]

It is clear that the slope of this line for fixed $n=0$~m$^{-2}$ is a ratio of the capacitances of the two gates. Similarly, the above equation can be solved for a fixed $D$, which means a fixed band gap, and only varying $n$:

\[V_{\rm tg}=\frac{C_{\rm bg}}{C_{\rm tg}}  V_{\rm bg}-\frac{C_{\rm bg}}{C_{\rm tg}}  V_{\rm b0}+V_{\rm t0}+2D\epsilon_0/C_{\rm tg}\]

The ($V_{\rm bg}$, $V_{\rm tg}$) co-ordinates corresponding to three electric fields $D = 0, -0.45$ and $-0.6$~V/nm are plotted in Fig.~S2b as black lines. Our measurements are performed along such straight lines where both $V_{\rm bg}$ and $V_{\rm tg}$ are varied simultaneously, maintaining a constant electric field (thereby, band gap) and varying carrier density.

\subsection{Probability distribution functions}

Fig.~S2 shows conductance probability distribution functions (PDFs) for several channel lengths at various (high) electric fields. Fig.~S2a shows the PDF at high carrier density and $|D|=0.75$~V/nm for Dev3, where the $\langle g \rangle \approx 8.5$. Clearly, this PDF can be best fitted with a Gaussian distribution. The variance $\langle (\Delta g)^2 \rangle$ is of the order of $(e^2/h)^2$ as expected from universal conductance fluctuations, in this weakly disordered metallic regime in BLG. In the same device, when the Fermi level is tuned to the CNP, the PDF shows a log-normal distribution which is made more clear in the insets. This is true for almost all the PDFs that we studied as shown in the Fig.~S2b - o, as well as in Fig.~2 in the main text. In some of them (Fig. S2, panels f, h, i, j, m), we find more than one peak that can be fitted with a log-normal distribution of conductance. We have chosen the main or most dominant peak in these panels to compute the statistical properties of $\ln g_{cnp}$, with the expectation that they represent the conductance of the weakest link in the 1D hopping channel. The additional peaks in these cases probably represent other weak links in the path as suggested by the observation that their corresponding values of $R_{\ln g}$ do not deviate far from that of the main peak, as shown in Fig. S4.

The electric field $|D|$ required to realize strong localization may vary between devices. We observed in Dev15 ($L=6$~$\upmu$m, $W= 3.1$~$\upmu$m) that the log-normal distribution emerges at 1.3~V/nm. In Dev17 ($L=19.5$~$\upmu$m, $W=3$~$\upmu$m) even at 1.1~V/nm the conductance distribution is normal and the device is in a weakly localized regime (see Fig.~S4). The corresponding conductance distributions are shown in Fig.~S3a - d. We believe, this may be because in a wider device ($\sim$3~$\upmu$m) there is a larger possibility of bulk channels to shunt the source and drain. Hence, only at larger fields all bulk states gap out and we realize the strongly localized regime. (These bulk channels may arise from local potential fluctuations, or a network of grain boundaries as shown in Ref.~\cite{Ju2015}).

%In spite of the appearance of multiple peaks in some of the cases, which is possibly due to mesoscopic inhomogeneity, the full statistical evaluation of the distribution function gives very similar trend in $R_{\ln g}$ as that obtained from Gaussian fits to the distribution function.

In some of the measurements, the PDF of $\ln g$ can exhibit weak asymmetry due to blocking effect or ``optimal shunts or punctures'' between viable localized sites in long and short channels, respectively~\cite{Raikh1989,Hughes1996}. We also have observed this in a few of our results, as shown in Fig.~S3e and f, where we have fitted with the RR model in which the distribution function is given by~\cite{Raikh1989}

\begin{equation}
P(\Delta)=\frac{e^{\Delta}}{\pi}\int_{0}^{\infty}dx\,\text{exp}\left(-x^{\nu^{1/2}}\text{cos}\frac{\pi\nu^{1/2}}{2}\right)\times\text{cos}\left(xe^{\Delta}-x^{\nu^{1/2}}\text{sin}\frac{\pi\nu^{1/2}}{2}\right)
\end{equation}
\label{RR}

where, 

\[\Delta=-\text{ln}\,g-\nu^{1/2}T_{o}/T\]
\[\nu=\frac{2T}{T_{o}}\text{ln}\left(\frac{L\nu^{1/2}}{\xi}\right)\]

\begin{figure}
\includegraphics[width=1\linewidth]{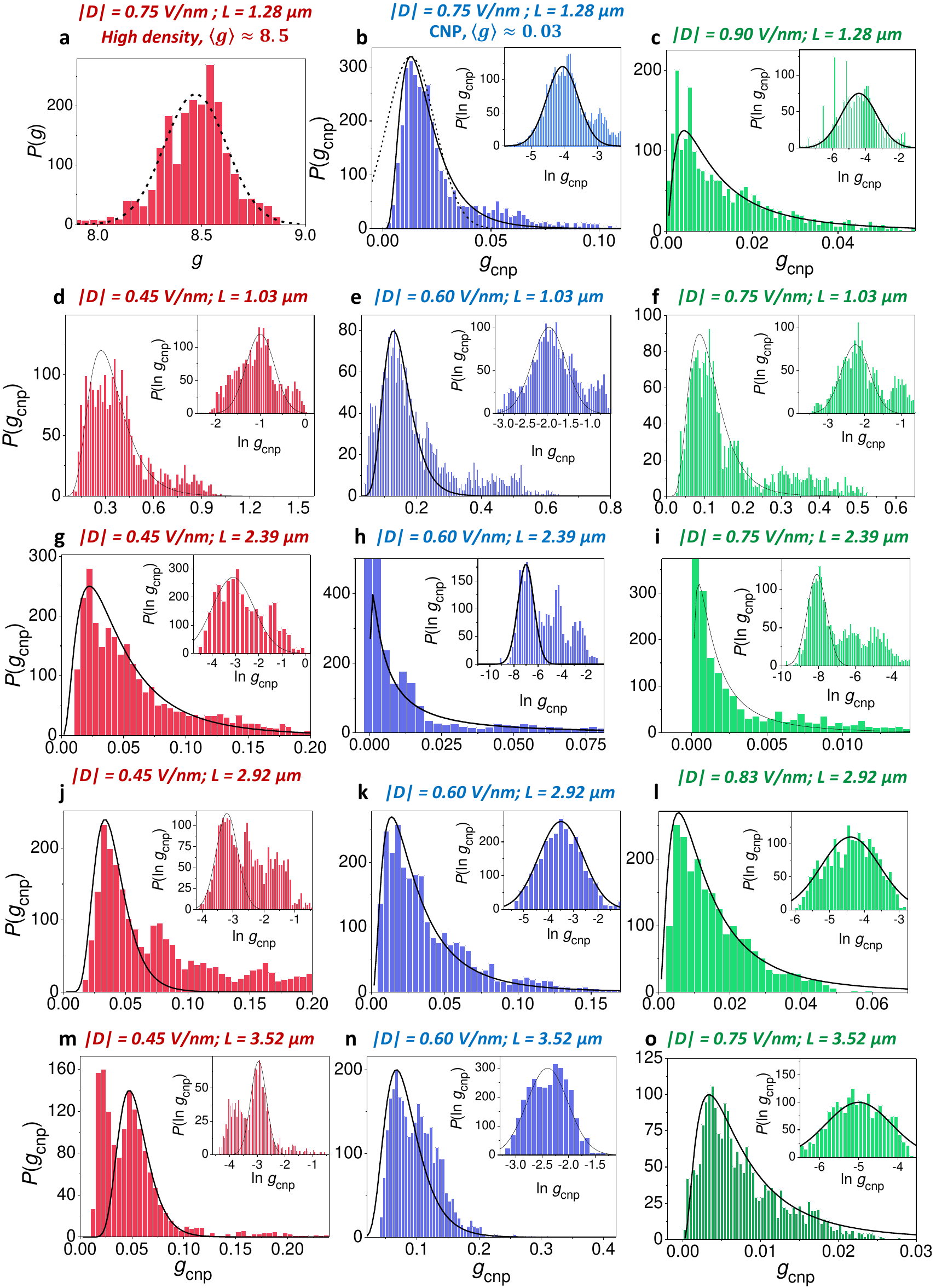}
\caption{PDFs of several devices for various electric fields}
\end{figure}
\label{FigS2}

\begin{figure}
\includegraphics[width=0.7\linewidth]{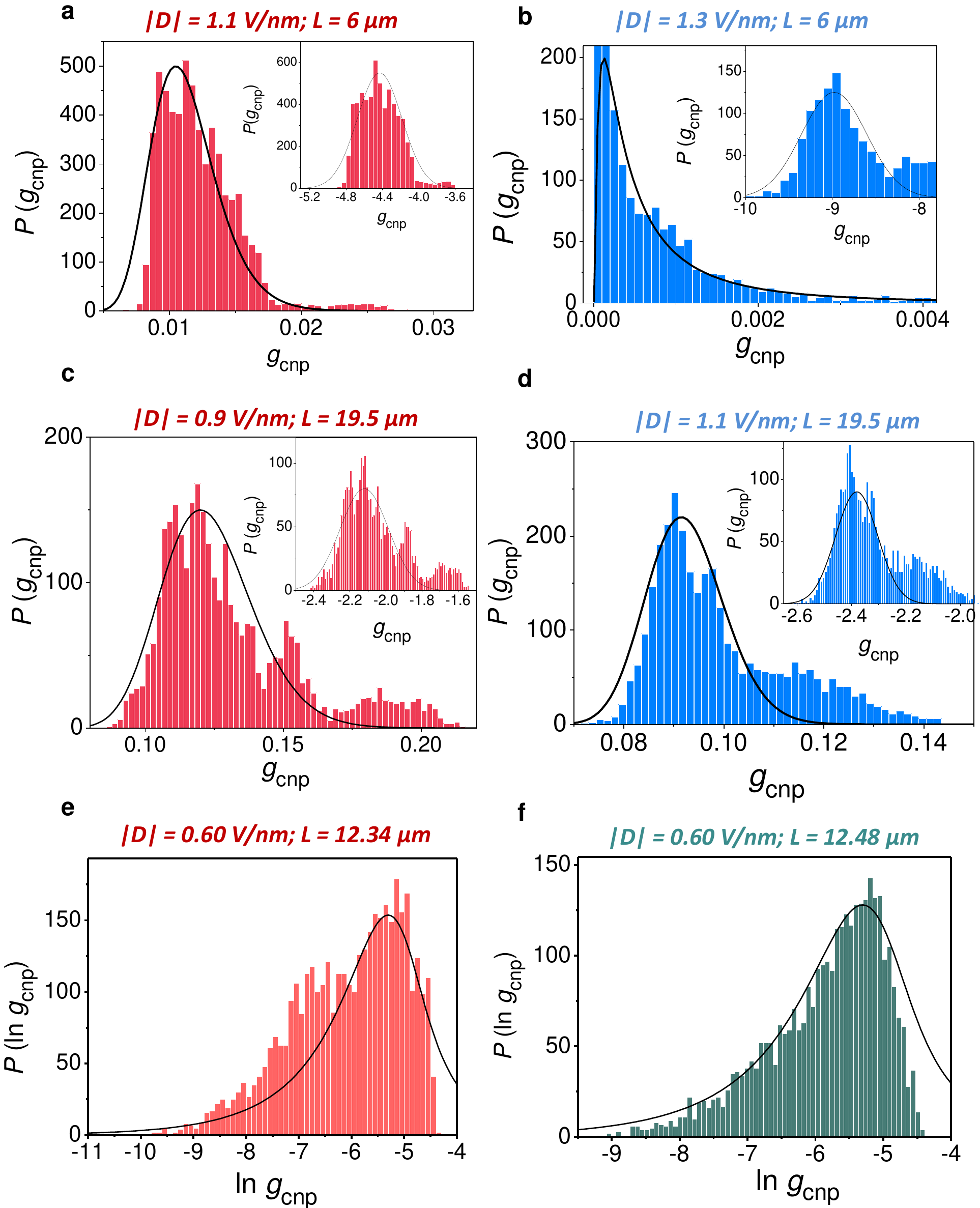}
\caption{\textbf{a, b.} Emergence of log-normal distribution at higher electric fields of $|D|=1.3$~V/nm for channel length 6~$\upmu$m. \textbf{c, d.} Log-normal distribution does not emerge unequivocally upto $|D|=1.1$~V/nm for channel length 19.5~$\upmu$m. \textbf{e, f.} Asymmetric PDFs which have been fitted with the RR model as given by eqn.~(1).}
\end{figure}
\label{FigS3}

\newpage

\begin{figure}
\includegraphics[width=1\linewidth]{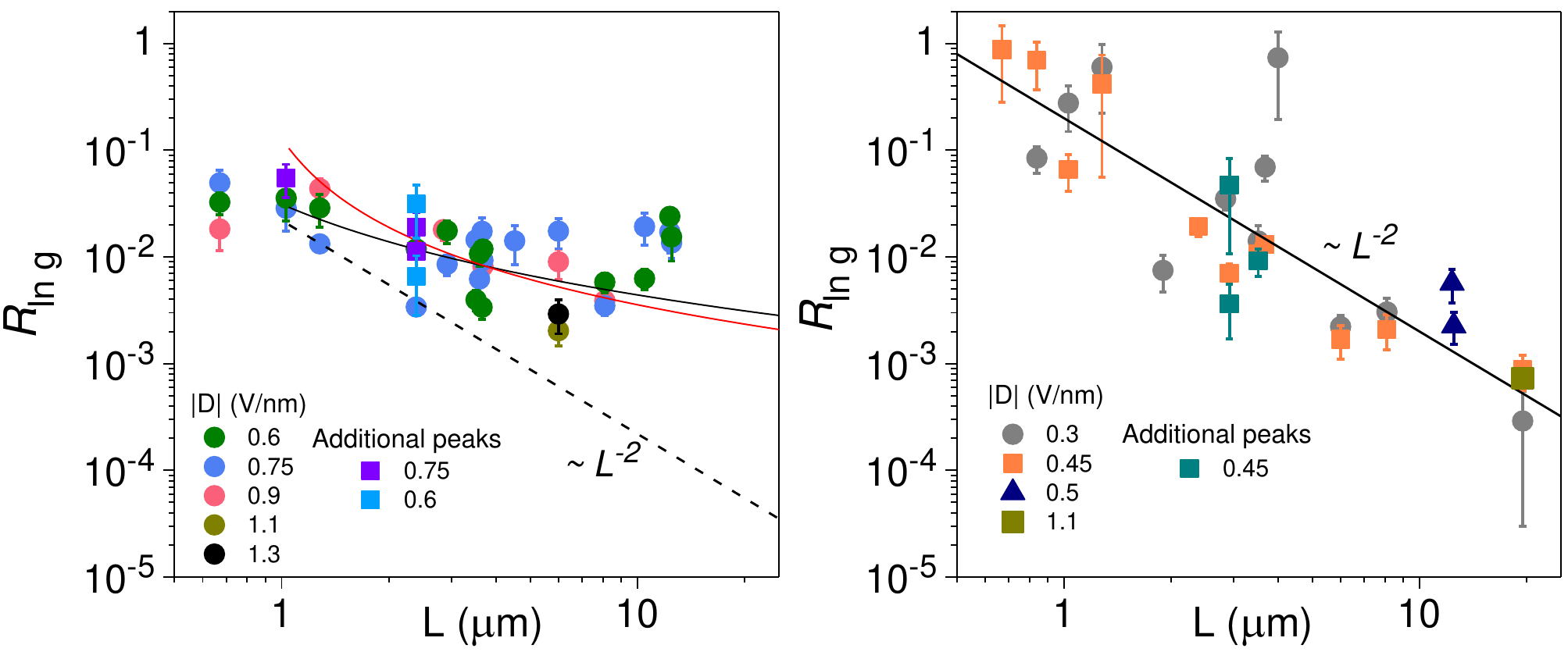}
\caption{Reproduction of Fig. 3a from the main text showing $R_{\ln g}$ as a function of $L$, now including evaluation for the occasional additional peaks in the conductance distribution.}
\end{figure}
\label{FigS4}

\newpage

\newpage

\subsection{Variance of $\ln g$ as a function of its mean}

\begin{figure*}
\includegraphics[width=0.75\linewidth]{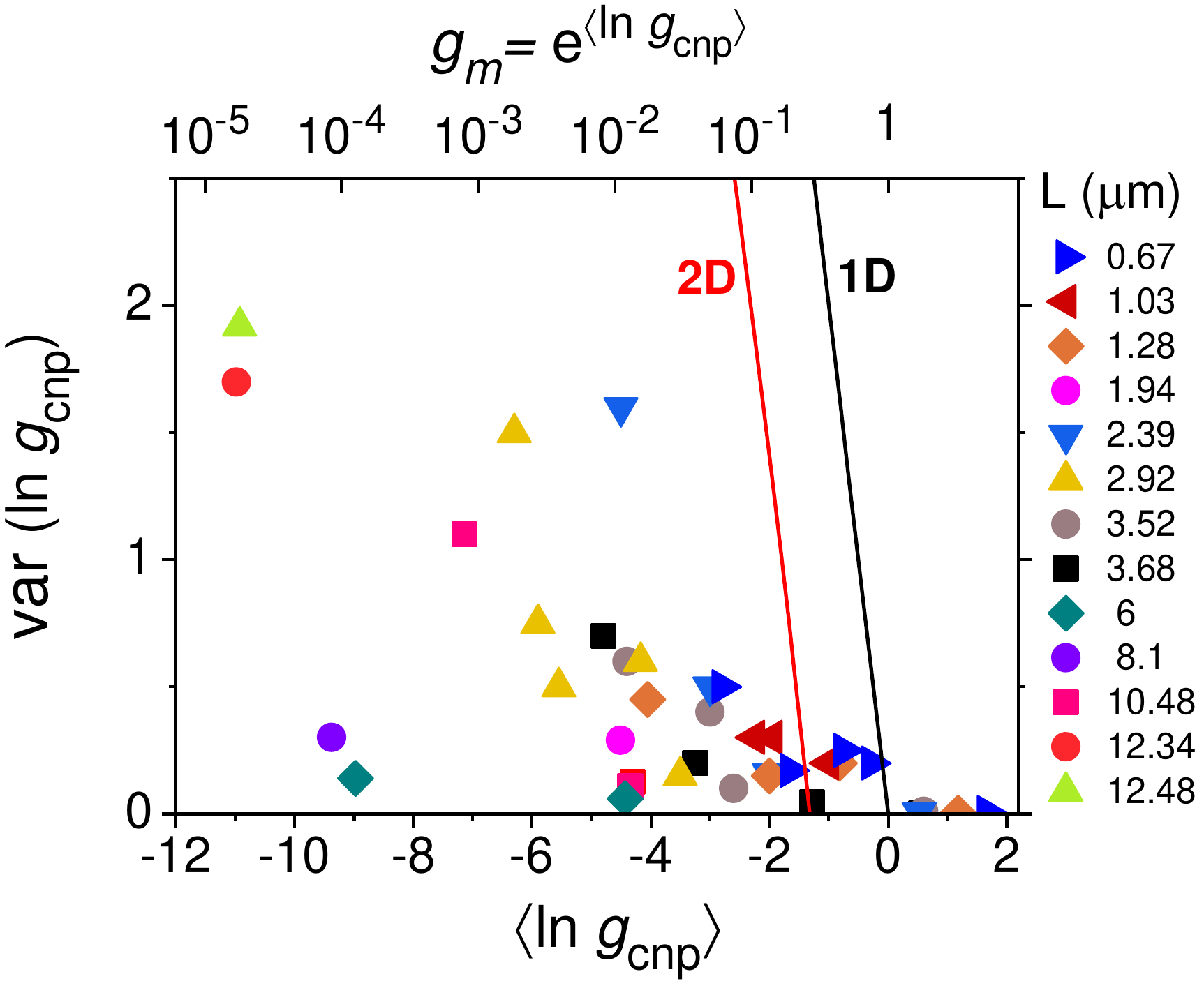}
\caption{Plot of variance ${\rm var}(\ln g_{{\rm cnp}})$ as a function of mean $\langle \ln g_{\rm cnp} \rangle$ as obtained by fitting of the PDFs for all $|D|$ ranging from 0 to 1.3~V/nm obtained by fitting $P(\ln g_{\rm cnp})$ for each $|D|$. They have been compared with the predicted behaviour of 1D~\cite{Beenakker1997} and 2D~\cite{Somoza2007}}\end{figure*}
\label{FigS5}

We have plotted the variance ${\rm var}(\ln g_{\rm cnp})$ as a function of the mean $\langle \ln g_{\rm cnp} \rangle$ in Fig.~S5. These parameters are extracted from the normal distribution fits to the PDFs. We note that the values of the variance are much lower than numerically predicted for both 1D~\cite{Beenakker1997} and 2D~\cite{Somoza2007} disordered systems at $T=0$~K. This further implies the invalidity of the conventional Anderson-like localization, supporting our arguments in the main text. However, at the same time, we also note that traces of ${\rm var}(\ln g_{\rm cnp})$ vs $\langle \ln g_{\rm cnp} \rangle$ seem to follow a trend for most of the devices, thereby implying that ${\rm var}(\ln g)$ may be following an approximately universal dependence on $\langle \ln g_{\rm cnp} \rangle$ with a few exceptions. Thus, it seems that $\langle \ln g_{\rm cnp} \rangle$ may be sufficient in describing the normal distribution of $\ln g_{\rm cnp}$ for most devices. Therefore, the single-parameter scaling hypothesis~\cite{Kramer1993} may hold in the localization in gapped BLG, statistically speaking, even though Anderson localization is not valid.

%In the second method, we have computed $var(\ln g)$ and $\langle \ln g \rangle$ from individual traces like the ones shown in the right panels of Fig.~1d in the main text. This has been plotted in Fig.~S6b for several channel lengths. We observe that the curves converge towards a similar point as the localization becomes stronger ($\langle \ln g \rangle\rightarrow -\infty$). Furthermore,  as the channel length increases, the curves also cluster together, as if approaching a universal curve. Again, this route of analysis also seems to imply the applicability of single-parameter scaling hypothesis.

%In Fig.~S6b, we have also plotted the variance and mean as obtained by fitting the log-normal distributions $P(\ln g)$. $\langle \delta(\ln g)^2 \rangle$ from all the different channels and $\Delta$ seems to be following a single parameter scaling, upto a margin highlighted by the shaded region bounded by the dashed lines.

%Such remarkably low values of variance indicate that there might be some novel mechanisms of transport in bilayer graphene. One possibility is that this transport is aided by quasi-ballistic 1D channels in the edges of the device~\cite{Li2010} or along grain boundaries~\cite{2015} that become important in the strongly gapped regime. We have attempted to resolve this by studying a theoretical model with localized states scattered across the sample and with some of them connected by such 1D channels. <Description, results and conclusions of the theoretical modelling>.

\clearpage

\subsection{Current-Voltage characteristics}

\begin{figure*}
\includegraphics[width=1\linewidth]{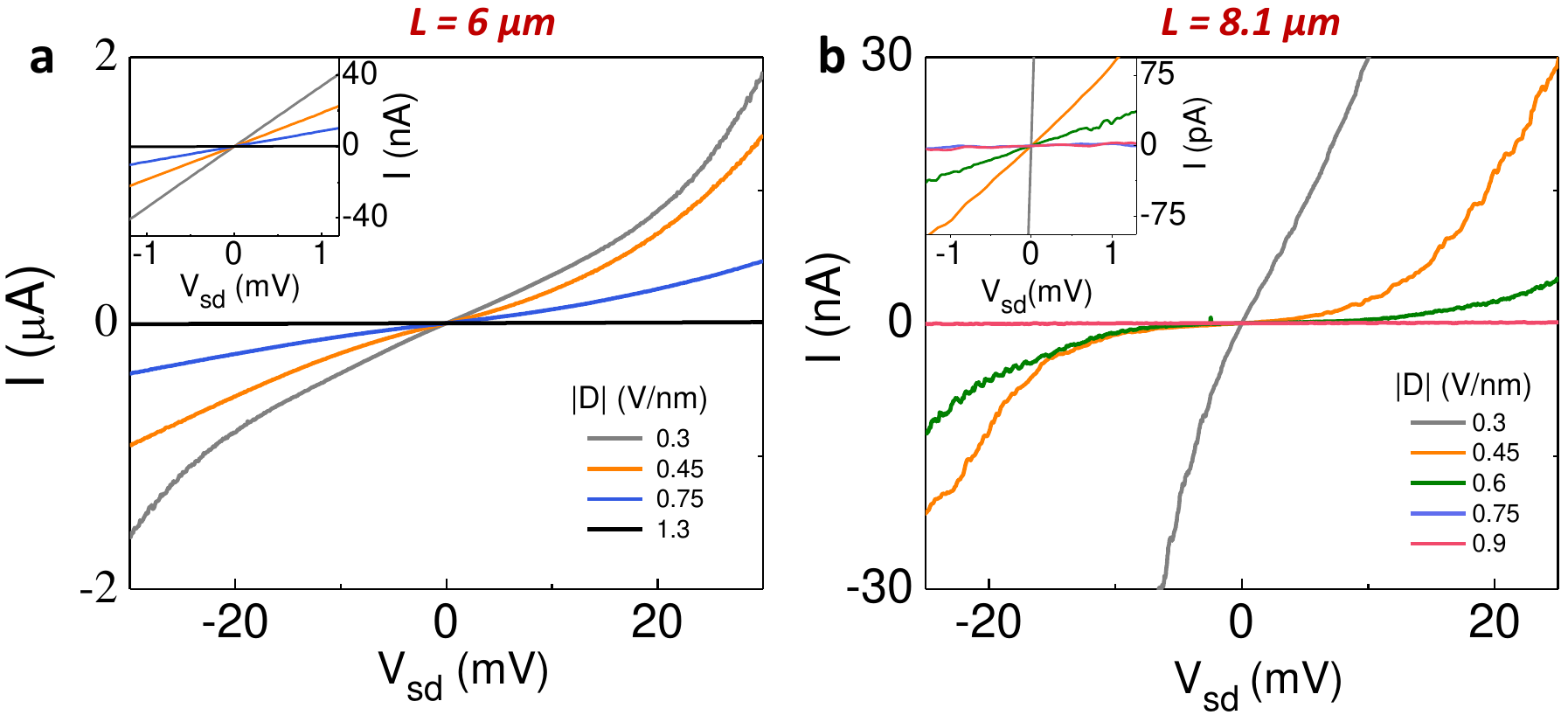}
\caption{$I-V_{\rm sd}$ characteristics of \textbf{a.} 6 $\upmu$m channel (Dev15)  at $T=4.2$~K and \textbf{b.} 8.1 $\upmu$m channel (Dev16) at $T=100$~mK, for $D$'s at the CNP. The insets show magnified $I-V_{\rm sd}$ characteristics, within a range of 1~mV in $V_{\rm sd}$.}\end{figure*}	
\label{FigS6}

$I-V_{\rm sd}$ characteristics of two devices (Dev15 and Dev16) are shown in Fig.~S6. They were measured by performing $dI/dV_{\rm sd}$ measurements by AC+DC adder method, followed by numerical integration. We observe a clear non-linear behavior, which is consistent with the presence of an effective transport gap. However, the onset of non-linearity occurs only on a scale of a few mV. Below 1~mV (insets of Fig.~S6), the $I-V_{\rm sd}$ characteristics are linear for all electric fields.

All conductance measurements in this work were carried out with source-drain bias (AC only) ranging from 10 $\upmu$V to 400 $\upmu$V and therefore, within the linear regime. The higher biases were applied in order to improve signal to noise ratio in the measurement of extremely low conductance.

\subsection{Author contributions}

M. A. A., P. K. and A. J. contributed equally to this work. M. A. A., P. K. and A. G. designed the experiment. M. A. A., P. K., A. J. and T. P. S. fabricated the devices. M. A. A., P. K. and A. J. carried out the measurements, and along with A. G. performed the analysis. T. V. R. and R. S. provided the theoretical input. M. A. A., P. K. and A. G. wrote the manuscript with input from all authors.